\documentclass[twocolumn,rmp,floatfix]{revtex4}
\usepackage{psfig}

\newcommand{\upd}{{\rm d}}
\newcommand{\kT}{k_{\rm B}T}

\begin{document}

\title{Initiation and dynamics of hemifusion in lipid bilayers}
\author{Guy Hed}
\author{S. A. Safran}
\affiliation{Department of Materials and Interfaces, Weizmann Institute of
             Science, Rehovot 76100, Israel}

\begin{abstract}
One approach to the understanding of fusion in cells and model
membranes involves stalk  formation and expansion of the hemifusion diaphragm.
We predict theoretically the initiation of hemifusion by stalk expansion and
the dynamics of mesoscopic hemifusion diaphragm expansion
in the light of recent experiments and theory that suggested that
hemifusion is driven by intra-membrane tension far from the fusion zone.
Our predictions include a square root scaling of the hemifusion zone
size on time as well as an estimate of the minimal tension for initiation of
hemifusion. While a minimal amount of pressure is evidently needed for stalk
formation, it is not necessarily required for stalk expansion.
The energy required for tension induced fusion is much smaller than that
required for pressure driven fusion.
\end{abstract}

\maketitle

\section{Introduction}
\begin{figure*}[t]
\centerline{\psfig{figure=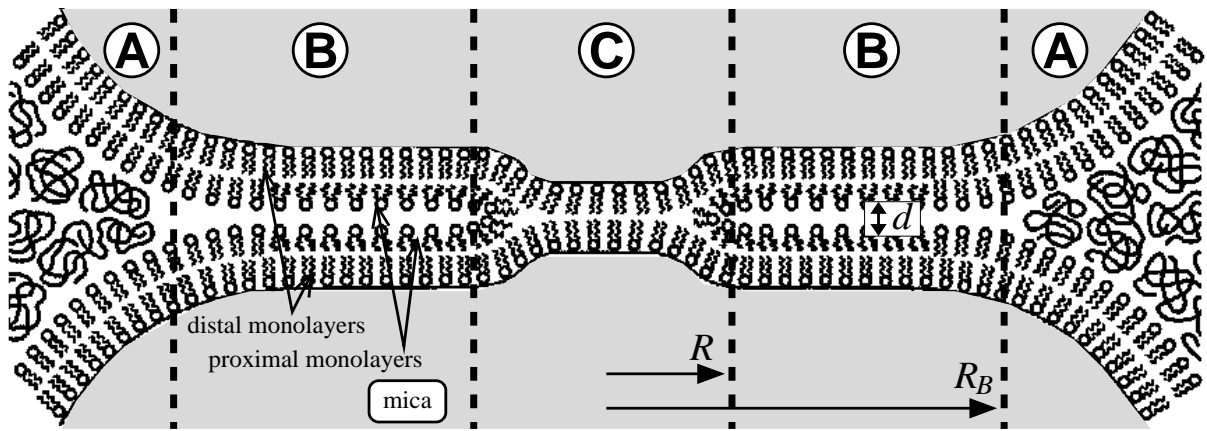,width=16cm}}
\caption{Illustration of the experimental geometry, adopted from
\textcite{kuhl96}. Regions A, B and C are defined in the text. $R$ and
$R_B$ are the inner and outer radii of region B, respectively.}
\label{fig:setup}
\end{figure*}

Membrane hemifusion is a possible pathway (see \textcite{muller02} for
an alternative view) to the complete fusion of membranes
\cite{chernomordik95}. Current theories associate the initiation
of hemifusion with the formation of a contact zone between the
membranes in which the two proximal monolayers are connected by a
stalk-shaped neck. The stalk then expands and a region is formed
(region C in Fig. \ref{fig:setup}), in which the two distal
monolayers form a single bilayer.
In general, the energetic cost of the splay of the lipid chains in
the stalk, prohibits its spontaneous expansion. However, the
presence of additional, external forces (e.g. pressure, surface
tension gradients, electrostatic effects) can lead to expansion of
the stalk into a 'hemifusion region' and to the growth of this zone.
Clear evidence for the existence of these two distinct
pre-fusion stages, stalk formation and hemifusion, was found
for PEG mediated fusion of vesicles \cite{lee97}.

A recent theoretical paper \cite{safran01} suggested that the flow of
lipids from region B to region A can be caused by an increase of
the surface tension in region A due to the presence (in that
region only) of additional polymer in solution. The tension
gradient between these regions induces a flow of lipids, that
leads to the growth of region C.

A different scenario, where hemifusion can be
an alternative pathway to fusion was
found in influenza hemagglutinin-mediated fusion \cite{chernomordik98,leikina00}.
The initial local stalk may evolve to a fusion pore \cite{muller02},
or it may expand to hemifusion. In the latter case, no fusion occurs.

In this paper, we predict the dynamics of the expansion of the
initial stalk and its role in the growth of a mesoscopic
hemifusion diaphragm. The nucleation of a stalk by thermal fluctuations
was recently shown to be thermally accessible
\cite{kozlovsky02,markin02}.  A detailed description of the
kinetics of this nucleation event (that typically describes the
formation of a stalk of several nanometers in extent)
is outside the scope of our work. Instead, we focus on estimates
of the conditions that facilitate stalk expansion into hemifusion.
We discuss the implications of our
theory on biological fusion mechanisms and on in-vitro experiments.
In addition, we predict the growth of the hemifusion region
(e.g. from nanometers to microns) as a function of
time and discuss the physical parameters that can be used to
control the time scale for hemifusion. This dynamic part is relevant
mainly to in-vitro experiments, since biological fusion events
generally remain at the microscopic scale of the stalk.

If hemifusion is an intermediate state of fusion then it is important
to contrast the time scales of hemifusion diaphragm
expansion and pore formation, in order to determine
the rate limiting step. \textcite{chizmadzhev00} predicted that pore
expansion is exponential in time, with a time scale of
$\eta_m/\delta p < 1$ sec, 
where $\eta_m$ is the membrane viscosity and $\delta p$ is 
the surface tension difference (both are estimated below). However,
if pore nucleation is slow enough significant expansion of the hemifusion
diaphragm can occur before pore formation. This is the case considered here,
where we predict that the hemifusion diaphragm expands as the square root
of time.

Our theoretical model is motivated by and consistent with the experiments
described by \textcite{kuhl96}, where two bilayers supported on mica surfaces
were brought into contact in the presence of a
PEG-water solution. Hemifusion, that eventually extended over a
distance of $50 \mu$ was observed in a time of about 10 minutes,
while the time it took the initial stalk to form was less then 3
minutes. This suggest that, at least in this experiment, the rate
limiting step for hemifusion
is the expansion of the fusion zone, as opposed to stalk formation.

This paper presents a simple
theoretical model relevant to this experimental system
\cite{kuhl96}, and predicts the time dependence of hemifusion
expansion. The overall time scale we find is comparable with the
measurements of \textcite{kuhl96} while the details of the
predicted temporal dependence have yet to be tested experimentally.

\section{Physical model}
\label{sec:thmod} Our theoretical model is illustrated in Fig.
\ref{fig:setup} that is a simplification of the experimental
system of \textcite{kuhl96} wherein two bilayers deposited on mica
cylinders are brought together in a solution of PEG and water. The
lipids of the distal monolayers are physisorbed on the mica; this
fixes their lateral density. From here on in this paper, the term
'lipid density' relates to the lateral density of the proximal
monolayers (see fig. \ref{fig:setup}).

We assume that the lipids are in {\em local} equilibrium, so at a
particular location $\vec r$, the free energy per lipid (in the
proximal monolayers) $\mu(\vec r)$, does not depend on the lipid
microstate, but only on the lipid density $\sigma(\vec r)$. This
assumption of local equilibrium is consistent with our results
that predict an overall time scale for hemifusion expansion that
is much larger than the local diffusion time of a single lipid
molecule.

The experimental system we consider is macroscopically
cylindrically symmetric and we therefore assume cylindrical
symmetry of all the physical quantities at mesoscopic length
scales. This is justified because all flows (of water and lipids)
are laminar, and there are no mechanisms that might induce angular
fluctuations or instabilities.

We distinguish between three regions, illustrated in Fig. \ref{fig:setup}:
\begin{itemize}
\item {\bf Region A} - where the distance $d$ between the bilayers is
typically much larger than the polymer correlation length $\xi$
\cite{safran01}. In this region, the outer lipid monolayer is in
contact with the PEG in the solution. The free energy per molecule
in this region is given by $\bar\mu(\sigma(\vec r))$, and is
different (in its functional form) from the free energy $\mu(\sigma)$
of the monolayer in the absence of PEG.

\item {\bf Region B} - where $d\ll\xi$. For
these values of $d$, the PEG density near the bilayers is
negligible and our model assumes that there is no PEG in contact
with the bilayers in this region. The free energy per lipid in
this zone is $\mu(\sigma(\vec r))$. In addition we assume that the
distance between the mica surfaces is constant (the mica
surfaces in the experiment are deformed and flattened under
pressure), and that this region is ring-shaped with an outer
radius $R_B$, and an inner radius $R$.

\item {\bf Region C} - the region where the distal bilayers are in contact.
\end{itemize}

The bilayers are Langmuir-Blodgett deposited in water, without PEG,
which is  added later.  The energy per lipid when the monolayers
are in contact with water is $\mu(\sigma)$ and the proximal
monolayers are initially Langmuir-Blodgett deposited
with a density $\sigma_0$ that minimizes
$\mu$. When PEG is added, it induces an effective attraction
between the polar heads \cite{safran01}, and changes the
functional form of the energy as function of the lipid density to
$\bar\mu(\sigma)$.

The effect of lipid condensation in the presence of PEG
\cite{tilcok79,batrucci96,maggio78} hasa been discussed in
terms of the the dehydration
of the bilayer by the PEG \cite{mishima97}. This dehydration affects
the lipids in region A that are in microscopic proximity to the PEG,
but has no effect on the lipids in region B. In Sec. \ref{sec:pressure}
we demonstrate that the osmotic pressure induced by the PEG is too small
to induce hemifusion. This stands in contrast to the surface tension effects
that are the main focus of our work.

If equilibrium could be reached, the lipid density in region A would tend to
increase in the presence of PEG. However, the number of lipids in the
monolayers cannot increase to any significant degree within the
time scale of the experiments, since the concentration of lipids
in the bulk solution is negligible and the number of lipids that can be
transported from region B to region A
is much smaller than that of region A. Thus, the lipid density is
unchanged and the energy per lipid in region A is now
$\bar\mu(\sigma_0)>\mu(\sigma_0)$, with the derivative $\bar\mu'(\sigma_0)<0$
due to the induced head attraction. This condensation effect  thus
leads to a {\em negative tension} 
in the proximal monolayers that
ideally would cause them to contract in extent.  They cannot do
this without exposing the chains of the inner monolayers to the
water and this is energetically prohibitive.  The outer monolayers
are therefore stressed and one way of relieving that stress is for
additional lipid to enter this region; this will allow the local
lipid density to increase, while still covering the original area
occupied by the outer monolayer.

The PEG concentration near the outer monolayers in region B is
given by $c_B=c_A (d / \xi)^2$, where $c_A$ is the PEG
concentration near the outer monolayers in region A, $\xi$ is
polymer correlation length, and $d$ is the distance between the
bilayers in region B \cite{safran01}. Since by the definition of
region B, the bilayer spacing in that region is small, $d\ll\xi$,
we have $c_B\ll c_A$ and the PEG concentration in region B is
negligible; we thus take this concentration to be zero. The energy
per lipid in region B is \emph{initially} given by
$\mu(\sigma_0)$, where $\sigma_0$ is the lipid density in the
absence of polymer. Since the free energy per lipid, $\mu$, is
minimized when the density $\sigma = \sigma_0$ and the tension
in region B initially vanishes, since either expansion or compression
of the lipids will increase their energy. The tension
gradient between regions A (initially at negative
tension) and B (initially at zero tension)
induces a flow of
lipids from region B to region A. Since region A is much larger
than region B, we can treat it as a reservoir, and assume that
even though lipid is flowing from region B to region A, the lipid
density in region A is not changed from its initial value of
$\sigma_0$.  The system is a dynamical one and the chemical
potential (equivalent in our single component system to the free
energy per lipid, $\mu$) is not constant in all of space at the
mesoscopic scale; this results in lipid flow and dynamics.
However, since local equilibrium \emph{is} maintained, we must
have equal chemical potentials at any given point in the system.
In particular, at the boundary between regions A and B, the
chemical potentials of the lipids must be equal:
$\mu(\sigma_b)=\bar\mu(\sigma_0)$, where $\sigma_b=\sigma(R_B)$ is
the lipid density at the edge of region B.  We note that this
equality of chemical potentials determines the lipid density at
the boundary of region B, $\sigma_b$; the functional form of the
two free energies $\mu(\sigma)$ and $\bar\mu(\sigma)$ are not the
same, since in region A, the lipids are in contact with polymer.

The initial lipid density in region B ($\sigma_0$, which is the
density at which the lipids self-assemble in water in the absence
of polymer) is higher than the lipid density at the AB boundary:
$\sigma_0>\sigma_b$.
This inequality is a consequence of the fact that the tension
at the boundary is negative, as shown in section
\ref{sec:boundcond}.
More intuitively, the negative tension in region A tends to pull
in additional lipids from the boundary region of region B into
region A as explained above.  This lipid flow reduces the lipid
density at the boundary $r=R_B$ from $\sigma_0$ to $\sigma_b$. In
turn, the reduced lipid density at the boundary of regions A and
B, ($\sigma_b < \sigma_0$) induces a flow of lipids from the rest
of region B towards the boundary.  This is because the minimum
energy state in region B is one where $\sigma=\sigma_0 >
\sigma_b$; thus lipids from the entirety of region B flow to the
boundary in an attempt to restore the lipid density there to
values closer to $\sigma_0$. This flow, in turn, reduces the lipid
density at the boundary between regions B and C (the hemifusion
region) at $r=R$, and lead to a negative tension 
that tends to expand region C.

At the boundary of regions B and C, the lipid density is
determined by a force balance between the membrane negative tension 
(arising from the lipids flowing to the AB boundary), that tends to expand
region C, and the force exerted by the boundary ring around region C
that tends to shrink it.
The main contribution to the energy of this ring is of
the tilt of the lipid tails imposed by the toroidal geometry.
This tilt is needed in order to form the three-way junction of the
boundary ring
cross section while avoiding an intra-membrane void, which has a much higher
energetic cost \cite{kozlovsky02}. The energetic cost of the tilt can
be considered through the related intra-membrane strain and the adjacent
stress tensor \cite{hamm00}.

We assume that for $R\gg d$ the energetic cost $f_t$ for a cross section of
the BC boundary ring is independent of $R$. Thus, the ring energy
is given by $E_r(R)=2\pi R f_t$. The force per unit length that
the ring exerts on region B of the membrane tends to shrink region C
and pull region B in the $-\hat r$ direction.  This force (per unit
length) is
\begin{equation}\label{eq:fbc}
\frac{1}{2\pi R} \left(- \frac{\partial E_r}{\partial R} \right) \hat r =
-\frac{f_t}{R} \hat r \;,
\end{equation}
and tends to shrink the boundary ring; that is, the
expansion of region C is energetically costly. In local
equilibrium, this force is balanced by the
surface tension ${p}$, which may be considered as a two dimensional
lateral lipid pressure ,
in region B of the monolayer that tends to expand the ring:
\begin{equation}\label{eq:fbeq}
{p}(R) + \frac{f_t}{R} = 0 \;.
\end{equation}
Negative tension 
in region B tends to cause this region to
contract and thus provides a force in the $\hat r$ direction,
balancing the force due to the BC boundary.

\begin{figure*}[t]
\centerline{\psfig{figure=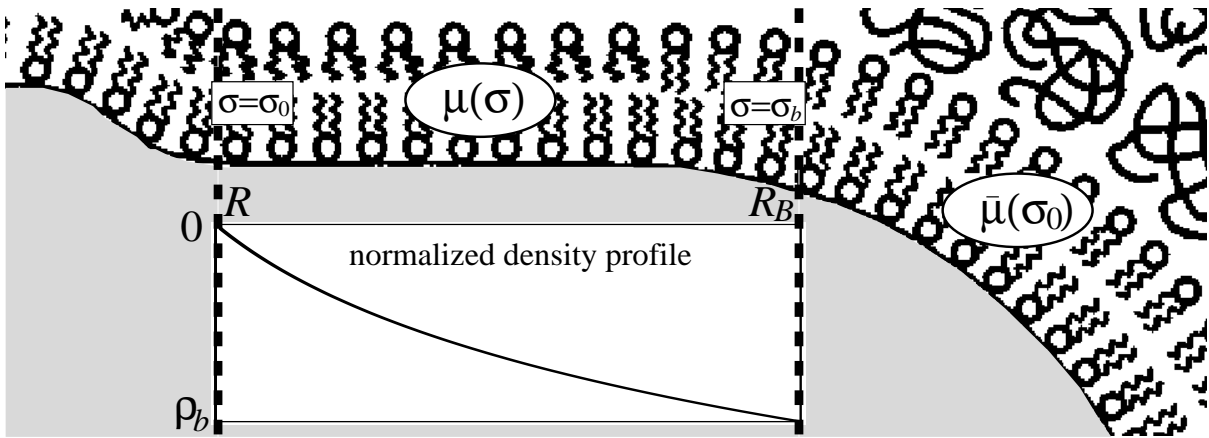,width=16cm}}
\caption{Enlargement of region B (between the dashed lines) shown
in Fig. \ref{fig:setup}. The lipid density in region A is the initial
density $\sigma_0$. In our model we assume a step in the density profile,
so the lipid density at $R_B$ is $\sigma_b<\sigma_0$.
At $R$ it is approximated by $\sigma_0$.
The normalized lipid density profile
$\rho=(\sigma-\sigma_0)/\sigma_0$ as a function of the radius $r$ for
$R/R_B=0.2$ is plotted using Eq. \ref{eq:lpdens}.
The free energy per lipid
in region B is $\mu(\sigma)$ which is a function of the local
lipid density. In region A, the free energy per lipid
$\bar\mu(\sigma_0)$ everywhere.}
\label{fig:monolayer}
\end{figure*}

\section{Monolayer dynamics}

In this section we  derive the dynamics that govern the expansion
of the hemifusion region and predict the flow of
lipids within the monolayer as a function of the lipid density and
of time.

There are three local, dissipative forces that oppose any lipid
motion.
\begin{itemize}
\item The stress, or force per unit area due to the viscosity of
the water that is moved along with the lipids is given by $\eta
\partial v_w/\partial z$, where $v_w$ is the water velocity and
$\eta=0.01$ erg s/cm$^3$ is the  viscosity of water. The stress is
of order $\eta v / d$, where $v$ is the lipid velocity and $d$ is
the spacing between the bilayers in region B. \item The stress, or
force per unit area due to the monolayer viscosity is given by
$\eta_m\nabla^2 v$, where $\eta_m$ is the monolayer friction
coefficient \cite{seifert93}. For a laminar flow we estimate this
stress as $\eta_m v / {R_B}^2$; that is, the relevant dimension is
the size of region B in which there is monolayer flow. \item The
stress, or force per unit area that is due to the friction
\emph{between} the monolayers is given by $b v$, where $b$ is the
friction coefficient.  This stress depends only on the motion of
the outer relative to the inner monolayer where there is no flow;
there is therefore no dependence of the length scale related to
the geometry of the different regions.
\end{itemize}

The friction between a DMPC monolayer and a supporting
HTS\footnote{trichlorosilanes with hexadecyl chains}
monolayer at $T=45^\circ \rm C$ is $b=7\cdot 10^6$erg s/cm$^4$,
while for a supported OTS\footnote{trichlorosilanes with octadecyl chains}
monolayer the friction is $b=2.9\cdot 10^8$ erg s/cm$^4$ \cite{merkel89}.
The experiments of \textcite{kuhl96} were carried out at $25^\circ \rm C$.
It has been observed
that the diffusion coefficient of a molecule in a DMPC monolayer
increases about three folds when $T$ is increased from
$25^\circ \rm C$ to $45^\circ \rm C$ \cite{vaz85,merkel89,haibel98},
which suggest a corresponding decrease in $b$. In this work we use
an estimated value of $b=10^8$ erg s/cm$^4$. For DMPC bilayers at
$T=25^\circ \rm C$ the bilayer viscosity is $\eta_m\sim 3\cdot
10^{-7}$ erg s/cm$^2$ \cite{merkel89}.
The values relevant to the experiments of \textcite{kuhl96} are $d =
2\cdot 10^{-7}$cm and $R_B = 5\cdot 10^{-3}$ cm. With the estimates
for the stress given above, we find that the frictional force due
to relative motion of the two monolayers is much larger than
either the lipid or water viscosity contributions to the stress.
We thus neglect these latter two effects and predict the dynamics
for a system where the only relevant dissipation is due to the
relative friction between the monolayers.

The lipid flow is induced by the tension gradient
$\nabla {p}$, and is  opposed by the frictional $bv$.  The force
balance equation is
\begin{equation}\label{eq:fb}
-\nabla {p} - b v =0 \;.
\end{equation}
In Appendix \ref{ap:local} we derive the lipid local dynamics using
Eq. \ref{eq:fb}
and the continuity equation. We consider the dynamics only to first
order in the lipid density variations $\rho=(\sigma-\sigma_0)/\sigma_0$,
which is known from experiments to be small. In \textcite{kuhl96} a
variation of $|\rho|\approx 0.05$ was measured.

To first order in $\rho$ the local dynamics has the form of
a diffusion equation
\begin{equation}\label{eq:eqmot}
\frac{\partial\rho}{\partial t}= \frac{\alpha}{b} \nabla^2\rho \;,
\end{equation}
where $\alpha={\sigma_0}^3 \mu''(\sigma_0)$ is the harmonic 'spring
constant' of the monolayer. For a small density
variation $|\rho|\ll 1$ the 
surface energy 
cost is $\delta g= \frac{1}{2}\alpha\rho^2$, and the related
tension difference is $\delta p=\alpha \rho$. 
We note that the {\em surface energy} $g=\sigma\mu(\sigma)$ is
the Gibbs free energy per unit area, and is different then
the {\em surface tension} $p$, which has the thermodynamic role of
the two dimensional pressure.

We estimate $\alpha$ using the phenomenological form
\begin{equation}\label{eq:musigma}
\mu(\sigma)=\gamma \left( \frac{1}{\sigma} + \frac{\sigma}{{\sigma_0}^2} \right)
\;,
\end{equation}
where $\gamma$ is the effective surface tension of the
hydrocarbon-water interface \cite{benshaul95}. The second term in
Eq. \ref{eq:musigma} accounts for the (electrostatic) effective
head-group repulsion, while the first term represents the
effective hydrocarbon-water repulsion. We note that this effective
repulsion is smaller than the repulsion of the bare
hydrocarbon-water interface, and has been estimated as
$\gamma\sim20$ erg/cm$^2$ \cite{israelachvili91}.

From Eq. \ref{eq:musigma} we obtain $\alpha=2\gamma \sim
50$ erg/cm$^2$. For $b=10^8$ erg s/cm$^4$ the effective 'diffusion
constant' is $5\cdot 10^{-7}$ cm$^2/$sec. This quantity is larger
than the actual, microscopic diffusion constant measured for free
liquid bilayers above the gel transition, that are of the order of
$10^{-8} - 10^{-7}$ cm$^2/$sec \cite{vaz85,haibel98,sonnleitner99}.
The Einstein relation is not applicable in our case, since the
flow (that happens to scale like diffusion)
of the lipids from the high to low density regions is not due to
the random motion of the molecules, but due to the tension 
gradient $\alpha\nabla\rho$. Indeed, for a characteristic
molecular area $a=10^{-14}$ cm$^2$ we find that the related
energy per molecule is $\alpha a \sim 10\kT$.

\section{Boundary conditions and global dynamics}
\label{sec:boundcond}

The boundary conditions for the lipid density were already
discussed in section \ref{sec:thmod} and we review them here for
convenience. The local tension 
equilibrium at the boundary with region A
determines the local lipid density $\sigma_b$ at $R_B$.
In Appendix \ref{ap:local} we show that the tension
in the monolayer is given by
\begin{equation}\label{eq:pl}
p = \sigma^2\mu'(\sigma) \;.
\end{equation}
Since the tension in region A is negative, from the tension 
equality at the boundary we see that ${p}(R_B) = {\sigma_b}^2
\mu'(\sigma_b)$
is negative. Moreover, because the
function $\mu(\sigma)$ has a minimum at $\sigma_0$ it is convex in
a neighborhood of $\sigma_0$. If $\sigma_b$ is in that
neighborhood, then the condition
$\mu'(\sigma_b)<0$ yields that $\sigma_b<\sigma_0$.

The boundary ring near the hemifusion region at $R$ exerts a
force that opposes hemifusion expansion; this is because the
boundary energy of the hemifusion region is increased as this
region grows. This force is  locally balanced by the negative tension 
in region B where lipids are flowing towards region A. As
lipids pass from region B to A the lipid density in region B
decreases; the tension 
in region B, and in particular near its boundary
with region C, becomes more negative and pulls on region C causing its
expansion.

The density of lipids in  region B at the boundary $R$ is
determined from the force balance Eq. \ref{eq:fbeq}.
Using Eq. \ref{eq:pl} we may write Eq. \ref{eq:fbeq} as
\begin{equation}\label{eq:bc1}
\rho(R)=-\frac{f_t}{\alpha R} \;.
\end{equation}

Before the flow begins, the initial lipid density in region B is
$\sigma_0$, which implies that $\rho=0$. For this value of the
lipid density there is zero tension in region B,  the stalk does
not expand and hemifusion does not develop.
Due to the tension gradient between region B and A, lipids  flow
out of region B and a negative tension is built up. If at a
certain time the lipid density at $r=R$ is low enough so that
Eq. \ref{eq:bc1} is satisfied, the stalk begins to expand.

After the flow of lipids is initiated, lipids are removed from
region B as they flow towards region A and the lipid density in
region B is lower than $\sigma_0$.  The lipid density in region B
cannot, however,  be smaller than the value of $\sigma_b$, because
when $\sigma=\sigma_b$ the free energies per lipid in regions A
and B are equal, and the flow stops. Thus, we require
$\sigma_0\geq\sigma\geq\sigma_b$ in all of region B if there is to be flow
and stalk expansion that leads to  hemifusion. At an early
time after the stalk formation, while the
stalk does not expand, the lipid density in all of
region B approaches the equilibrium density profile
$\sigma(r)=\sigma_b$.
Using Eq. \ref{eq:bc1}, the condition for the stalk to begin to
expand with a finite amount of time is:
\begin{equation}\label{eq:inicond}
-\rho_b > \frac{f_t}{\alpha R_0} \;,
\end{equation}
where $\rho_b=(\sigma_b-\sigma_0)/\sigma_0$ and $R_0$ is the
radius of the stalk. In our model, we consider the process
for $R$ much larger than the molecular size $R_0$ that
characterizes the size of the stalk.
The tilt energy $f_t$ is in general
positive.  From Eqs. \ref{eq:pl} and \ref{eq:inicond}, for
$R\gg R_0$ we have $|\rho(R)|\ll|\rho_b|$. Since we consider all
quantities only to first order in $\rho_b$ we use the
approximation $\rho(R)=0$.

In Appendix \ref{ap:global} we use the integral continuity equation, that
expresses the conservation the lipid number in the system, in order to obtain
a dynamic equation for the hemifusion radius $R$. In Appendix \ref{ap:adiabatic}
we show that the time scale that governs the local dynamics is much faster
then the rate of change of $R$.
We use an adiabatic approximation in order to solve the
dynamics. First, we fix $R$ and find the asymptotic ($t\rightarrow\infty$)
lipid density profile
\begin{equation}\label{eq:lpdens}
\rho(r) = \rho_b \left( 1 - \frac{\log(r/R_B)}{\log(R/R_B)} \right) \;.
\end{equation}
We use this density profile to obtain the dependence of the hemifusion
radius $R$ on the time $t$ to find:
\begin{equation}\label{eq:tauR}
\frac{\alpha \rho_b}{b} t = R^2 \left(\log\frac{R}{R_B} -
\frac{1}{2} \right) \;.
\end{equation}
This predicts an approximately square root dependence of the hemifusion region
size on time (with logarithmic corrections). 
The same temporal dependence
was obtained by \textcite{kumenko99} under the assumption of constant
lateral lipid density. However, their result is quantitatively different
from ours since they have considered the monolayer viscosity as the main
dissipative force, while we have showed that it is negligible compared to
the friction $b$.

From Eq. \ref{eq:tauR} we find that the time it takes the
hemifusion region to evolve from the initial stalk radius
$R=R_0\ll R_B$ to a final radius of $R=R_B$ is $\Delta t = - b
{R_B}^2 / 2\alpha\rho_b$. With $\rho_b = -0.05$ and
$\alpha/b=5\cdot 10^{-7}$ cm$^2/$sec, we predict that the time for
expansion of the hemifusion zone to a scale of $R_B=50\ \mu$m is
$\Delta t\approx 500$ sec. This is consistent with the experiment of
\textcite{kuhl96} where a time of $\Delta t=600$ sec was measured.

The time $\Delta t$ found here can also be derived (up to a numerical
factor) from a simple scaling argument, that does not depend on the specific
details of our model. As hemifusion is initiated, the tension 
difference between the bulk (at $R_B$) and the hemifusion front (at $R$) is 
$-\alpha\rho_b$. When $R_B\gg R$ the average tension gradient is
$\overline{\nabla p}\approx -\alpha\rho_b/R_B$. For a fully damped flow
with a friction coefficient $b$ the average lipid velocity is
$\bar{v}=\overline{\nabla p}/b$. The hemifusion front (BC boundary)
advances with the velocity $\sim\bar{v}$ of the lipids near it.
The time to advance a distance of $R_B$ with a velocity $\bar{v}$ is
$R_B/\bar{v}= -b {R_B}^2 / \alpha\rho_b$.

\section{Initiation of hemifusion}
\label{sec:init}
The change in the monolayer surface energy due to the presence of
PEG in region A is $\delta g=\sigma_0(\bar\mu(\sigma_0) - \mu(\sigma_0))$,
where $\mu(\sigma_0)$ is the free energy per lipid in the absence of
PEG, and $\bar\mu(\sigma_0)$ is the free energy per lipid in the
presence of PEG. Since we have defined $\sigma_b$ by
the condition $\mu(\sigma_b) = \bar\mu(\sigma_0)$, we can expand
$\mu$ around its minimal value $\sigma = \sigma_0$, and find
that to lowest order in $\rho_b$ the surface energy difference $\delta g$
and the tension difference $\delta p$ induced by the PEG are
\begin{equation}\label{eq:tendiff}
\delta g = \frac{1}{2} \alpha {\rho_b}^2 \;;\ \ \
\delta p = \alpha \rho_b \;.
\end{equation}
In \textcite{kuhl96} a change of $\rho_b\approx -0.05$ in lipid
density was deduced from the measured thinning of the bilayer.
Using the value $\alpha=50$ erg/cm$^2$ we estimate $\delta g\approx
0.06$ erg/cm$^2$; $\delta p\approx 2.5$ erg/cm$^2$.

Initiation of stalk expansion is relevant not only to events of mesoscopic
fusion, but also to in-vivo fusion events, where a fusion pore is
formed soon after stalk expansion. In many cases of biological interest,
the fusion process is
regulated by fusion proteins that promote stalk formation
and expansion. One hypothesized bio-molecular mechanism that promote
expansion is the penetration of hydrophobic fusion protein domains into the
membrane and its subsequent destabilization \cite{bentz00}.
The protein domains may increase the membrane surface
energy by inducing an effective attraction of the hydrophobic head groups,
similar to the effect of PEG \cite{safran01}; they may also penetrate the
membrane, increasing the intra-membrane tension. 
Our theory suggests
that the former mechanism, which work to increase in $\delta g$, may be more
effective energetically than the latter, which increases $\delta p$.
That is, for a given change in lipid density, $\rho_b$, a smaller energy
is involved (Eq. \ref{eq:tendiff}).

SNARE\footnote{soluble N-ethylmaleimide-sensitive factor-attachment
protein receptors} proteins that promote exocytosis
in nerve synapses are thought to induce stalk expansion through a
conformational change by which the
protein pull on the stalk to widen it \cite{scales01}. Another possible cause
for stalk expansion is calcium ions induced membrane tension \cite{arnold95}.
We conclude from our theory that the latter mechanism may be more effective
energetically.

In section \ref{sec:boundcond} we found that in order for
expansion of the hemifusion region to occur, the driving force due
to the negative tension in region B must be large enough to
overcome the tendency of the boundary of region C to shrink.  We
thus deduced that the normalized lipid density at $R_B$ must obey
\begin{equation}\label{eq:rhob}
-\rho_b > \frac{f_t}{\alpha R_0} \;.
\end{equation}
From this condition, we estimate the minimum
stalk radius $R_0$ for which the lateral tension in the monolayer can
induce expansion. The energy of the lipid tails tilt at the
hemifusion front is estimated by \textcite{markin02} as
$f_t = 2\cdot 10^{-6}$ erg/cm. For the values of $\alpha$ and
$\rho_b$ given above, we find that the mechanism described here is
sufficient to cause hemifusion for $R_0 \geq 8$nm, which is of the
order of the typical radius of a thermally nucleated stalk \cite{yang02}. 
Note that if $\rho_b$ vanishes (that is, no polymer is present in
region A) hemifusion will not be initiated for
any finite (reasonable) stalk radius.

\section{The role of pressure}
\label{sec:pressure}
Experiments have demonstrated that hemifusion may be caused by sufficiently
large normal pressure \cite{helm89} or by negative pressure
in the water layer \cite{yang02,macdonald85}.
We shall now determine the
conditions under which pressure induced in region B
can in and of itself
(i.e. with no surface tension effects as induced by the added
polymer) cause hemifusion expansion by forcing water
to flow out of the contact zone.
We do this by using the simplifying assumption
that the water in region B is under a constant pressure
$p_w=p_n+p_o$, where $p_n$ is the normal pressure on the bilayers
and $p_o$ is the osmotic pressure induced by the solute in the bulk. 
The finite thickness of the water layer in region B (whose thickness is on the
order of a nanometer) is always maintained
because of hydration forces: the water molecules are organized
around the polar head groups of the lipids in order to partially cancel their
electric dipole; removing the water layer would increase the free energy
because of the energetic cost of these electric dipoles whose normal
components, in general, point to the same direction due to the hydrophobic
nature of the lipid layer.
Thus the water flow out of region B and into region A is possible only by the
expansion of region C.

The energy (per unit area) difference associated with a pressure
difference of $p_w$ is $p_w d$, where $d$ is the distance between
the two proximal monolayers. This should be compared with the
energy difference $\delta g$ associated with the free energy
gradient in the monolayer. In the experiment of \textcite{kuhl96}
that yield $p_w d \approx 0.08$ erg/cm$^2$, which is of the same order
of $\delta g$. Nevertheless, we  show below that the external normal
pressure has only a minor effect on the pressure in the monolayer and on
its density. We will thus show that under the experimental conditions of
\textcite{kuhl96}, the external pressure
is \emph{insufficient} to cause hemifusion expansion.

In the experiment of \textcite{kuhl96} the applied normal pressure is
$p_n=0.3$ atm and the osmotic pressure is $p_o \sim 0.1$ atm, so the
total pressure between the bilayers is $p_w \approx 0.4$ atm. 
We now estimate the contribution of this pressure to the lipid density
variation in the experiment.
For a fluid membrane, the relation between the tension ${p}$ 
- the two dimensional pressure in the membrane -  to the three
dimensional pressure $p_w$, is $p_w={p}/h$, 
where $h$ is the thickness of the monolayer.
In order to induce the observed density variation $\rho_b=0.05$ the tension 
needed is $|{p}|=2.5$ erg/cm$^2$. For $h=5$ nm the pressure required to 
induce such tension is 5 atm -- much larger than the actual pressure in
the experiment. Thus, the contribution of the normal and the osmotic pressures
to the density variation is negligible compared with the surface tension
effects due to the PEG-lipid interactions that result in densification of 
the lipids. This result underscores the point
made in Sec. \ref{sec:init}: changes in the pressure are much less effective
than surface energy variation for the initiation of stalk expansion.

We now estimate the pressure $p_w$ needed to initiate hemifusion,
without a lipid density gradient (that is, with $\rho_b=0$). The
radial force per unit length on the boundary at $R$ due to the
external normal pressure is
\begin{equation}\label{eq:fpress}
\frac{-1}{2\pi R}\frac{\partial (p_w V)}{\partial R}=p_w d \;.
\end{equation}
From Eq. \ref{eq:fbc}, the condition for spontaneous fusion is
$p_w d > f_t/R_0$. For the values given above, we require $p_w
\geq 10^7$dyne/cm$^2 =10$ atm. Experimental results in different
conditions are within that range. The pressure needed for the
hemifusion of bilayers directly supported on mica (with no added
polymer or other mechanisms that give rise to lipid density
gradients) was found by \textcite{helm89} to be $p_w \sim 40$ atm.
\textcite{wong99} used a surface forces apparatus to
apply pressure on DMPC bilayers supported on polymer layers. The
polymer layer allowed the bilayers some lateral conformational
freedom, thus permitting more freedom for the adjustment of stalk
shape and size \cite{kozlovsky02,markin02}. In that case, where the stalk
geometry could easily adjust, the cost for forming the stalk was
reduced and hemifusion was observed at a much lower pressure of
$p_w=2$ atm.
In the experiment of \textcite{kuhl96} the pressure $p_w\approx 0.4$
atm is too low to be the driving force for hemifusion.

Pressure in itself is not enough to cause hemifusion, but it is
sometimes necessary\footnote{In the experiments it is difficult to
distinguish between applied pressure and time in contact effects
(T. Kuhl, private communication).}. 
\textcite{leckband93} showed that the amount of
pressure needed for hemifusion is directly related to the lipid
density near the contact area. In that experiment, two bilayers
were brought into contact using a surface forces apparatus.
When Ca$^{++}$ ions were
introduced, there was a phase separation in the bilayers. The
density of lipids in the bilayer regions that were brought into
contact was characterized by the hydrophobic adhesion energy. When
thinner regions were brought together (characterized by adhesion
energy of $E_{\rm ad}=3.8$ erg/cm$^2$) they  either hemifused
spontaneously, or required only a small amount of pressure ($p_n
\leq 1$ atm) to induce hemifusion. For denser bilayers ($E_{\rm
ad}=0.15$ erg/cm$^2$) a pressure of $p_n=4$ atm was required for
hemifusion. 

\textcite{yang02} induced negative osmotic pressure on the water layer between
the bilayers by lowering the relative humidity
of the environment of DPhPC\footnote{diphaytanoyl phosphaditylcholine}.
At 80\% humidity the lipids were at the lamellar
phase. As the relative humidity was decreased the water were expelled from
between the bilayers by the osmotic pressure and the lamella were connected
by stalks, directly observed by x-ray diffraction. In this experiment the
dehydration was due to negative pressure of the water layer induced by
the reduced relative humidity, and not by normal pressure, but the physical
effect of the two is similar.

\section{Summary}

In this paper we used a model based on lipid density gradients
induced by surface energy variation 
that occur far from the hemifusion zone, to predict the the conditions
for the initiation of hemifusion by stalk expansion and the dynamics of
mesoscopic hemifusion. Our theory  was motivated by the
experiments of \textcite{kuhl96}. However, the
quantitative scheme presented here can be generalized to any
system of two lipid bilayers initially connected by a stalk, where
a perturbation in region A, mesoscopically far from the stalk, causes
tension in the membrane in that region. For example one can apply
our results to tension induced by the electrostatic interactions
caused by calcium ions \cite{leckband93}, tension induced by
laser tweezers \cite{barziv94,moroz96}, or the effective tension induced
by the attraction of oppositely charged bilayers \cite{pantazatos99}.

We have compared the effect of the friction of the two monolayers, the
the water viscosity and the intra-monolayer viscosity on the two dimensional
lipid motion and showed that
the friction dominates. Thus, the lipid dynamics depend on the friction
and not on hydrodynamics. This means that the spacing between the two layers
is irrelevant for the lipid dynamics.

Experiments similar to those of \textcite{kuhl96} can test the predictions of
the model for the time scales as
functions of the lipid density and friction as well as the value
of the driving force due to the tension induced in region A.
One could vary each of the parameters $\rho_b$ (the
relative change in lipid density), $\alpha$ (related to the
induced tension) and $b$ (the interlayer friction) independently,
and measure the 'hemifusion radius' $R(t)$, the final radius $R_B$
and the time to complete the process $\Delta t$ as  functions of
these parameters.

In particular, the
friction $b$ can be varied independently of $\alpha$ by changing
the composition of the distal bilayers while maintaining the same
composition of the proximal bilayers.
The friction can be varied by changing the interactions
between the chains that is responsible for most of the friction, via chain
length changes or temperature changes \cite{yoshizawa93}.

Once an empirical, temporal profile for the hemifusion expansion,
$R(t)$, is measured for systems with known parameters, one can use
the same experiment to estimate the effective diffusion constant
for the lipid flow, $\alpha/b$, for \emph{different} lipid
bilayers. One can easily vary the lipid density at the boundary,
$\sigma_b$ by changing the polymer (or calcium ions) concentration
since the density $\sigma_b$ is determined by the equality of the
chemical potentials of the lipids exposed to the polymer and those
exposed only to the water.

The static part of our theory deals with the initial conditions
required for stalk expansion. We have evaluated the necessary density
variation $\rho_b=-f_t/\alpha R_0$ and demonstrated that the
related surface energy 
$\frac{1}{2}\alpha{\rho_b}^2$ is much smaller then the surface tension
$\alpha\rho_b$. This result is not surprising, since it is
a general result of a first order expansion around an energetic minimum.
Still, it does give a new insight regarding biological fusion
mechanisms. It suggest mechanisms working
through the change of the surface energy $\delta g$ 
are much more effective than mechanisms that exert force or
normal pressure on the stalk.

The predicted dependence of stalk expansion on the lipid density can be
tested by measuring the critical density $\rho_b$
at which stalk expansion occurs. The results may serve to learn more
about the stalk structure and energetics.

We expect that near  the end of the process of hemifusion
expansion, when $R(t)\approx R_B$ experimental results may
differ from our predictions, since the density profile of the
polymer (or calcium ions in the case of \textcite{leckband93}) may
vary in a gradual manner around $R_B$; in our theory we assumed a
sharp ('step function') decrease of the polymer density at $R_B$.
We also expect a deviation from our theory  when the radius $R(t)$
of the hemifusion region is close to its initial, molecular stalk
radius $R_0$, due to microscopic details of the lipid structure in
the stalk.

We distinguish between hemifusion induced by surface tension gradients,
which we consider in our model,
and hemifusion induced by pressure. Hemifusion may be induced by normal pressure
on the bilayers \cite{helm89,wong99} or by dehydration which induces negative
pressure in the water layer between them \cite{yang02}. We showed that
this pathway to hemifusion requires much more energy (per unit area)
than fusion that is induced by surface tension gradients.

We have shown that the induced pressure $p_w$ 
in the experiment of \textcite{kuhl96}
cannot be the primary direct cause of hemifusion. Still, pressure does play an
important role in stalk formation. 
It may also effect stalk expansion through its effect
on the lipid tilt energy $f_t$ and on the initial stalk radius $R_0$.

\appendix

\section{Local lipid dynamics}
\label{ap:local}
We bring here the full calculation of the local lipid dynamics.
Note that though in our final result we leave only the terms linear in
$\rho$, one may also calculate in the same framework the non linear terms
in the case $\rho$ is not small.

The force balance equation is
\begin{equation}\label{eq:a_fb}
-\nabla {p} - b v =0 \;,
\end{equation}
and the continuity equation is
\begin{equation}\label{eq:a_cont}
\frac{\partial \sigma}{\partial t} + \nabla(\sigma v) =0 \;.
\end{equation}
Writing the energy per lipid as $\mu(\sigma)$  the surface tension is 
\begin{equation}\label{eq:a_pl}
{p}= -\left.\frac{\partial \left(N\mu(\sigma)\right)}{\partial
A}\right|_{N} = \sigma^2\mu'(\sigma) \;,
\end{equation}
where $A$ is the a macroscopic area and $N=\sigma A$ is the number of lipids
in this area.

From Eqs. \ref{eq:a_fb}, \ref{eq:a_cont} and \ref{eq:a_pl} we have
\begin{eqnarray}\label{eq:a_lpdyn}
b\frac{\partial\sigma}{\partial t}= \nabla(\sigma \nabla {p})=
\left( 2\sigma^2\mu'(\sigma)+\sigma^3\mu''(\sigma) \right)
\nabla^2\sigma + \nonumber\\
\left( 4\sigma\mu'(\sigma)+5\sigma^2\mu''(\sigma)+\sigma^3\mu'''(\sigma)
\right) (\nabla\sigma)^2 \;.
\end{eqnarray}
To first order in the density variation $\rho$,
Eq. \ref{eq:a_lpdyn} has the form
\begin{equation}\label{eq:a_eqmot}
\frac{\partial\rho}{\partial t}= \frac{\alpha}{b} \nabla^2\rho
+{\mathrm O}(\rho^2) \;,
\end{equation}
where $\alpha={\sigma_0}^3 \mu''(\sigma_0)$.

\section{Global lipid dynamics}
\label{ap:global}
In Sec. \ref{sec:boundcond} we consider the boundary conditions for the
lipid density.
In order to fully predict the dynamics of hemifusion
expansion, we also need to determine the flow at the
boundaries. For this we use the integral form of the continuity
equation:
\begin{equation}\label{eq:a_bc2}
\frac{\partial}{\partial t} \int_{R}^{R_B} 2\pi r\upd r \,\sigma(r) =
- \oint_{R_B} \sigma \vec{v}\cdot\upd\vec{l} \;.
\end{equation}
The left hand side of Eq. \ref{eq:a_bc2} describes the rate of
change of the lipid number in region B while the right hand side
gives the flow of lipids through the boundary $R_B$. We assume
cylindrical  symmetry, so $\vec{v}=v_r\hat{r}$. From Eqs.
\ref{eq:a_fb} and \ref{eq:a_pl} we obtain
\begin{equation}\label{eq:a_vrR0}
v_r(r)=-\frac{1}{b}\left( 2\mu'(\sigma) +\sigma\mu''(\sigma) \right)\sigma
\frac{\partial \sigma }{ \partial r} \;.
\end{equation}

We now use Eq. \ref{eq:a_eqmot} to calculate the left hand side of
Eq. \ref{eq:a_bc2}:
\begin{eqnarray}\label{eq:a_lhd1}
&& \!\!\!\!\!\!\!\!\!\!\!\!
\frac{\partial}{\partial t} \int_{R}^{R_B} 2\pi r\upd r \,\sigma(r) =\nonumber\\
&& \!\!\!\!\!\!\!\!\!\!\!\!
2\pi \frac{\alpha}{b}
\left( R_B\left.\frac{\partial\sigma}{\partial r}\right|_{R_B}
- R\left.\frac{\partial\sigma}{\partial r}\right|_{R} \right) -
2\pi  R \sigma(R) \frac{\partial R}{\partial t} \;.
\end{eqnarray}
If we take only terms linear in $\rho$, Eq. \ref{eq:a_bc2} gives:
\begin{equation}\label{eq:a_bc3}
\frac{\partial R}{\partial t} =
\frac{\alpha}{b} \left( \frac{R_B}{2R} \left.\frac{\partial\rho}{\partial r}
\right|_{R_B} - \left.\frac{\partial\rho}{\partial r}\right|_{R} \right) \;.
\end{equation}

\section{Adiabatic solution}
\label{ap:adiabatic}
Equations \ref{eq:a_bc3} and \ref{eq:a_eqmot} along with the boundary
conditions completely determine the time evolution of the
monolayers to first order in $\rho$. From these equations we can
calculate $R(t)$, and predict the temporal profile of hemifusion expansion.
We write these two equations using
dimensionless variables and scale the spatial variables so that
they are of order of unity, in order to get an estimate of the
time scales. The natural spatial scale is the final
size of the hemifusion region, $R_B$.  We thus define: $x=r/R_B$,
$\bar R=R/R_B$, $\bar \rho=\rho/\rho_b$ as well as two time
variables: a ``fast'' time $\tau = \alpha t / b {R_B}^2$ at which
the local lipid flow occurs, and a ``slow'' (since $\rho_b$ is
small) time $\bar\tau=|\rho_b|\tau$ which is the scale over which
the hemifusion region expands.
Eqs. \ref{eq:a_bc3} and \ref{eq:a_eqmot} become
\begin{equation}\label{eq:a_nsc1}
\frac{\partial\bar\rho}{\partial\tau}=\frac{\partial^2\bar\rho}{\partial x^2}+
\frac{1}{x} \frac{\partial\bar\rho}{\partial x} \;,
\end{equation}
\begin{equation}\label{eq:a_nsc2}
\frac{\partial\bar R}{\partial\bar\tau}=\left.
\frac{\partial\bar\rho}{\partial x} \right|_{x = \bar R} -
\frac{1}{2\bar R} \left. \frac{\partial\bar\rho}{\partial x} \right|_{x=1} \;.
\end{equation}

Since all the variables that appear on the right hand side of
Eqs. \ref{eq:a_nsc1} and \ref{eq:a_nsc2} are of order unity, the
units of $\tau$ and $\bar\tau$ suggest the time scales of the
processes described by the equations. For $|\rho_b| \ll 1$ we have
$\tau\gg\bar\tau$, which implies that we can use an adiabatic
approximation: the local lipid flow occurs quickly so that the
lipid density is instantaneously given by the asymptotic equilibrium solution
of  Eq. \ref{eq:a_nsc1} for $\tau\rightarrow\infty$.  We then use
this solution to determine the slower time evolution of the
hemifusion radius $R$ from Eq. \ref{eq:a_nsc2}.

At asymptotically long times, both sides of Eq. \ref{eq:a_nsc1}
vanish. The adiabatic density profile reached is
\begin{equation}\label{eq:a_lpdens}
\bar\rho(x) = 1 - \frac{\log(x) }{ \log(\bar R) } \;.
\end{equation}
Plugging this solution into Eq. \ref{eq:a_nsc2} we obtain
\begin{equation}
\frac{\partial\bar R}{\partial\bar\tau} = \frac{-1}{2\bar R \log(\bar R)} \;.
\end{equation}
The solution of this equation is  implicitly given by
\begin{equation}\label{eq:a_tauR}
2\bar\tau = {\bar R}^2(1-2\log\bar R) \;.
\end{equation}

\vspace{3em}
\begin{small}
We gratefully acknowledge useful discussions with Tonya Kuhl and
Jacob Israelachvili and the support of the Israel Science
Foundation and the Schmidt Minerva Center.
\end{small}




\end{document}